# Modelling and Measuring the Irrational behaviour of Agents in Financial Markets: Discovering the Psychological Soliton.

Gurjeet Dhesi

School of Business, London South Bank University,

103 Borough Road, London SE1 0AA, United Kingdom

Email : dhesig@lsbu.ac.uk (correspondence author's email)

Marcel Ausloos [1,2,3]

[1] School of Management, University of Leicester,

University Road, Leicester, LE1 7RH, United Kingdom

Email : ma683@le.ac.uk

[2] GRAPES , rue de la Belle Jardiniere 483,

B-4031, Angleur, Belgium

Email : marcel.ausloos@ulg.ac.be

[3] eHumanitiesgroup,

Royal Netherlands Academy of Arts and Sciences,

Joan Muyskenweg 25, 1096 CJ, Amsterdam, The Netherlands

Email : marcel.ausloos@ehumanities.knaw.nl

# Abstract

Following a Geometrical Brownian Motion extension into an Irrational Fractional Brownian Motion model, we re-examine agent behaviour reacting to time dependent news on the log-returns thereby modifying a financial market evolution. We specifically discuss the role of financial news or economic information positive or negative feedback of such irrational (or contrarian) agents upon the price evolution. We observe a kink-like effect reminiscent of soliton behaviour, suggesting how analysts' forecasts errors induce stock prices to adjust accordingly, thereby proposing a measure of the irrational force in a market.

**I. Introduction:**

A recent paper by Dhesi *et al.* [1] about modelling returns distributions for financial market indices, uses a quite innovative approach for obtaining a better fit. This is achieved by adding an extra stochastic function incorporating a weighting factor (with only two parameters to be estimated) over the well known Geometric Brownian Motion (GBM). This type of modelling, recalled in Section II, is endogenous and part of some coherent understanding of the market process, i.e. taking into account some so called irrationality of agents, sometimes accepted "by common knowledge" as a realistic possibility, but hardly included in models.

It can be recalled that Bulkley and Harris [2] showed that stock price volatility may be due (in part) to a failure of the market to form rational expectations, using data on analysts' expectations of long run earnings growth for individual companies. Simulations provided (on various data sets) in the Dhesi *et al.* paper [1] show that this model, for describing agent behaviours, lead to very good fits (tested by $\chi^2$ tests) to the empirical returns distributions of various empirical price indices. The fits proposed by Dhesi *et al.* [1] are far superior compared to those obtained through the ordinary GBM. In particular, the Dhesi *et al.* [1] model captures the fat tails and overall leptokurtosis. Therefore, it can be claimed that the model makes a fully pertinent connection between the extra function and irrational behaviour of financial markets. Further justification of this connection is provided in Section III below.

Thus, here, we re-examine the specific role of agent effects on the market following this GBM extension, for short let it be called IFBM (Irrational Fractional Brownian Motion) model. Rational expectations of investors with respect to future prices of assets/shares are assumed to reflect any available information. This is the essence of the efficient market hypothesis (EMH) [3,4].

Under the EMH, agents not following the GBM strategic investment advice are called irrational agents. This behavior is not recommended because it is usually "concluded" that the agents would loose money in the long run because of their irrationality, being way off any understanding on how the market evolves. Becker [5] contends that such agents do not (rationally) intend to maximize their profit (!), whence, in modern language, their utility function. Prechter [6] believes that such agents ride social mood waves, are following emotions, have an irrational behavior, … whence are irrational, - yet, they are rational without knowing so. Of course, agents can be switching between two trading behaviors, like informed vs. liquidity traders [7]. In fact, previously, De Long *et al.* [8] already analysed the impact of noise traders in the financial markets. Their findings pointed out that the so called irrational behaviors of traders are the main instigators of price volatility in the market, whence acting to the advantage of other groups of investors! Thus, it might be acknowledged that such so called irrational behaviors determine fluctuations in the movements of prices away from the fundamental/intrinsic value [9].

Nevertheless there is some evidence that these irrational agents might be able to obtain higher than average returns (see Brock and Hommes in [10, 11] for an appropriate discussion). To use "noise", fractional Brownian motion with a Hurst coefficient, $H \neq 0.5$, rather than the classical Brownian motion, has been demonstrated indeed to be very useful for investor strategies [12-16]. These strategies are "very rational", apparently paradoxical, but, on the contrary, much thought of. Some part of the paradox, should be distinguished from so called "irrational expectations", popularized through irrational exuberance [17].

To describe the irrational behavior of agents, a so called bounded rationally confident agent model is usually mimicked [18]. It implies an awareness threshold determining the level to which an agent puts a confidence weight. This necessarily "non-linear model" leads to features like avalanches, bubbles, etc. [19, 20]. Of course, we do not claim that these features are only (or even mainly) due to irrational agents nor to irrational behaviors.

It has been argued that markets are not inherently rational, "but are driven by fear and greed". Lo [21, 22] has argued that rational and irrational behaviors are "opposites sides of the same coin" and an evolutionary approach can be applied so as to reconcile "market efficiency with behavioral alternatives". Considering this, Lo [21] pointed out that labeling of investors who do not follow the EMH as irrational investors is inappropriate; a more accurate term for them should be "maladaptive". This provided some motivation to develop the adaptive wave alternative for option pricing model [23].

Bearing in mind that the markets are driven by fear, greed and impatience, a different way can be considered within a more direct stochastic (and to a large extent non adaptive) function as introduced in the GBM evolution equation for the share prices [1]. In the following, in Section II, we re-explain this new model and the new stochastic function theoretical origin. Section III interprets the financial return evolution and inherently explains the modelling of the irrational behaviour of the market. Section IV elucidates how the irrational behaviour of the market can be considered as a non-adaptive "psychological soliton" of the financial markets. Beside in so doing proposing a measure of the irrational force component in a market, further directions in related research are outlined.

## II. The model.

Of course, there can be a debate on what is to be considered as rational or irrational behaviour, as briefly outlined in Section I through a few references. In line with the Efficient Market Hypothesis, we attribute rational behaviour of the markets to the notion that the market price incorporates all information rationally and instantly. Thereafter, the irrational component has to be introduced.

Let us start with the usual financial log-returns definition

$$r_t = \ln(\frac{P_t}{P_{t-1}}) = \mu + \varepsilon_t \qquad \varepsilon_t : NID(0, \sigma^2) \qquad (1)$$

where μ is the average return and $\varepsilon_t$ is assumed to be normally independently distributed (NID) with zero mean and constant variance σ (a "white noise"), within the underlying assumptions of the error term in the EMH. The above equation can be written as

$$\ln(P_{t+\delta t} / P_t) = \mu \delta t + \sigma Z_t \sqrt{\delta t} \qquad (2)$$

in which $Z_t$ is a random number, drawn from a standardised normal ("Gaussian") distribution and $\delta t$ is a small time step. This equation is deployed to model returns distributions based on the Geometric Brownian Motion (GBM), e.g. see Peters [24] or Paul and Baschnagel [25]. However, the distribution of returns generated from this GBM model does not match the distributions of historic returns data which often show leptokurtosis [26, 27]. Motivated by an experimental paper due to Dhesi *et al.* [28], Dhesi *et al.* [1] added a function of the random number $Z_t$ weighted by the mean and an extra parameter $K$, in order to describe the returns distribution through

$$\ln(P_{t+\delta t} / P_t) = \mu \delta t + \sigma Z_t \sqrt{\delta t} + \mu K f(Z_t) \delta t \qquad (3)$$

thereby leading indeed to a much better fit to the log-return distributions, in particular in the peak and the tails: see [1]. This modified specification is important as this endogenously generates a distribution which is not arbitrarily exogenously imposed, but demands to choose the appropriate realisation of $f(Z)$, that is leptokurtic and hence is appropriate for the returns distributions.

Therefore, this modelling process suggests ways to describe irrational behaviour in finance: details are provided in section III. It should be noticed that only the normally distributed $Z_t$ innovation appears in the model, in contrast to e.g. the numerous models of returns distributions using jump diffusion processes which contain normal innovations and Poisson jumps [29].

The above discrete time evolution of log-returns can be transformed into a stochastic differential equation, by applying Ito's Lemma:

$$d \ln P = \alpha dt + \sigma Z \sqrt{dt} + \alpha K f(Z) dt \qquad (4)$$

valid when $\mu dt$ is small, and where $\alpha \equiv \mu + \frac{1}{2}\sigma^2$. Obviously for $K = 0$, the GBM is recovered.

After much extensive empirical analysis of historical data on various market indices, Dhesi *et al.* [1] have proposed that $f(Z) \equiv f_c(Z)$ is

$$f_c(Z) = (2 e^{(-c\frac{Z^2}{2})} - 1).\arctan(Z) \qquad (5)$$

To let the reader be aware of the role of the parameters *K and c* a few cases are illustrated on Figs. 1-3.

Fig. 1 Plot of the information feedback function $K f_c (Z_t)$ for $K = -5$, $c = 1$ as a function of $Z_t$

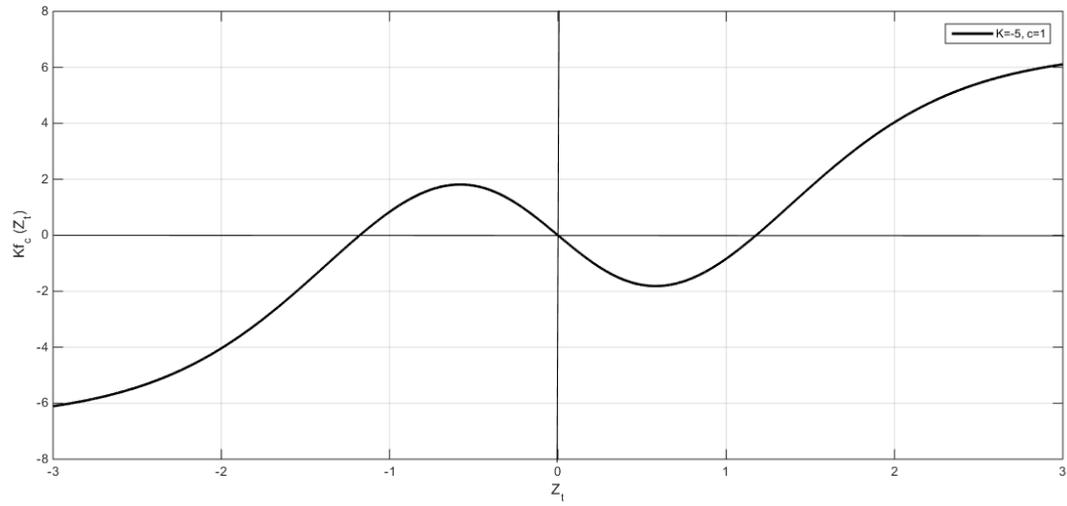

Fig. 2 Plot of the information feedback function $K f_c (Z_t)$ for various $K$ values $= -1, -5$, and $-10$, and for $c = 1$ as a function of $Z_t$

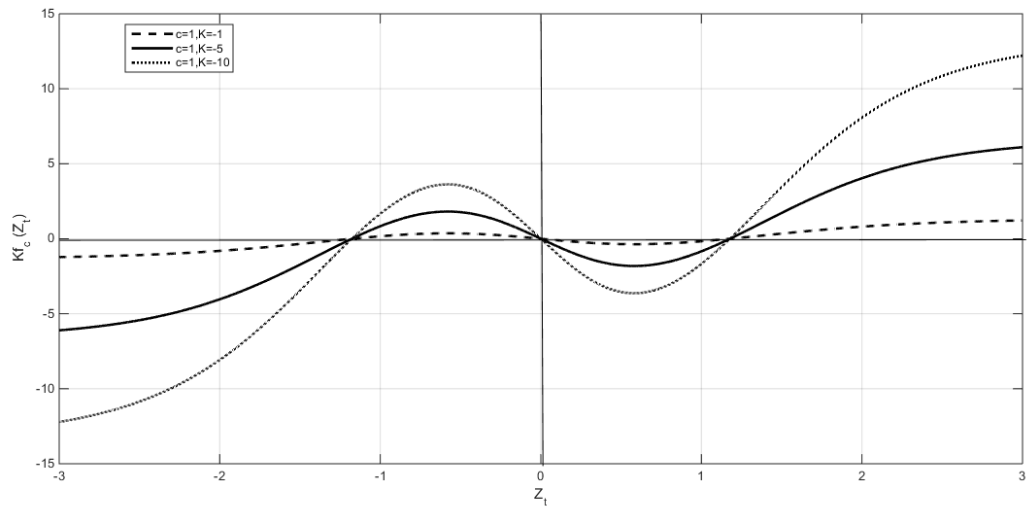

Fig. 3 Plot of the information feedback function $Kf_c(Z_t)$ for $K = -5$, and various $c$ values : $c = 0.5, 1,$ and $2,$ as a function of $Z_t$

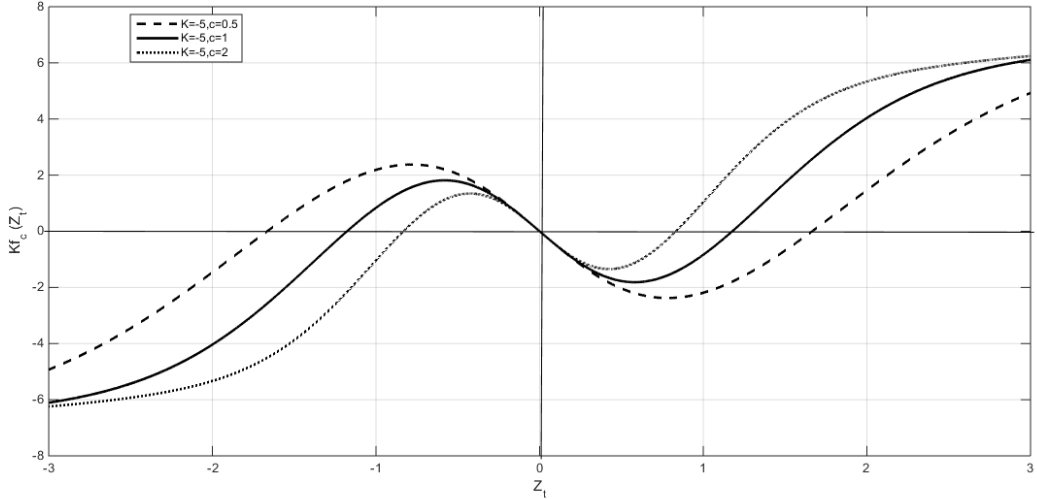

Fig. 1 allows to illustrate the overall shape of $Kf_c(Z_t)$ for $c=1$. The non trivial behaviour allows us to point to a change in curvature and slope near $Z_t = 0$, leading to a local minimum and a local maximum. Fig. 2 shows the influence of $K$ for a given $c$ : neither the roots move, nor the local extrema, but their amplitude increases with $K$. Fig. 3 shows how the extrema and roots move as a function of $c$. Any reader has observed that for $K$ negative, the function has its local minimum for $Z_t$ positive, and its local maximum for $Z_t$ negative. They are of equal magnitude in absolute value. Thus, it seems worth to point out here that the feedback of information function, when multiplied by $K$, is an "irrational feedback" function when $K < 0$, as further developed in Section III.

### III. Financial return evolution interpretation: measuring/modelling the irrational.

Consider for the sake of discussion that $Z_t$ corresponds to some financial news or economic information at time particular time $t$. The feeding parameter to the investor is $Kf_c(Z_t)$ it gathers and provides the additional output beyond the GBM at each time step $t$. Simulations are then performed using the antilog version of Eq. (3) and returns distribution is modelled though aggregation [1].

We will term $Kf_c(Z_t)$ as the feedback function. This feedback function for negative values of $K$ and arbitrarily for $c = 1$ for illustrative purposes is provided in the annotated figure 4. The following analysis of this section, and referring to Eq. (3), explains that $Kf_c(Z_t)$ measures and models the irrational feedback behaviour of agents when $K<0$.

Fig. 4 Plot of the information feedback function $Kf_c(Z_t)$ for $K = -5$ and $c = 1$ as a function of $Z_t$, emphasizing the various regions of interest.

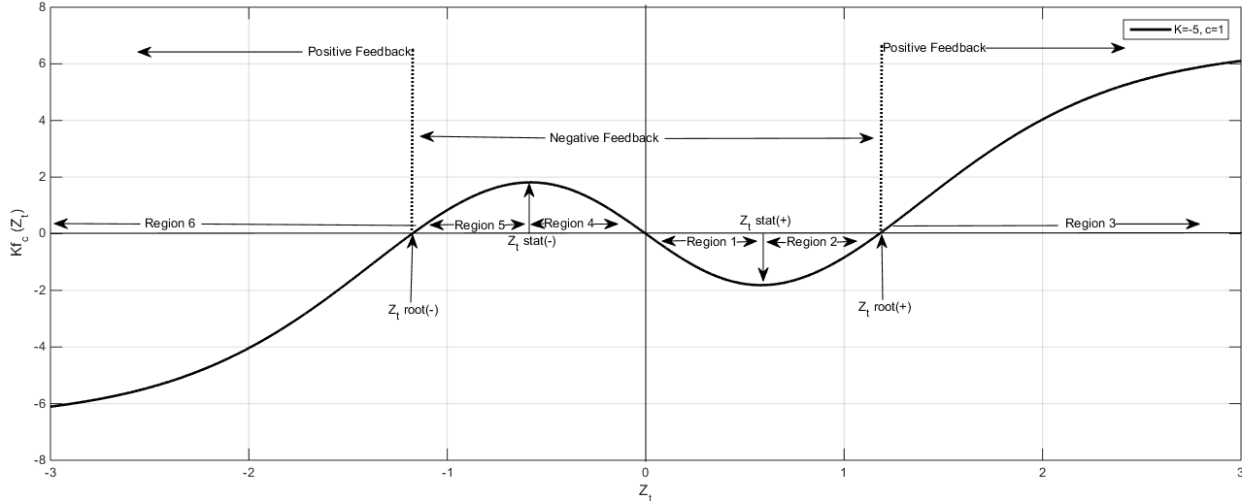

Figure 4 contains six regions separated by roots and extrema of $Kf_c(Z_t)$. Assuming that we are currently at time $t$, we will first consider the regions 1, 2 and 3 corresponding to positive values of $Z_t$.

Region 1: $0 < Z_t < Z_{t,stat(+)}$ the $\delta P_t$ contribution from GBM is positive, with the slope of $Kf_c(Z_t) < 0$; the $\delta P_t$ contribution/correction from $Kf_c(Z_t)$ is negative. This implies that $(P_{t+\delta t})_{IFBM} < (P_{t+\delta t})_{GBM}$. As $(P_{t+\delta t})_{IFBM}$ is closer to $P_t$, this peaks the returns distribution. This region is a negative feedback region as $Z_t Kf_c(Z_t)$ (product of $Z_t$ and $Kf_c(Z_t)$) is negative. The price rise in the IFBM is smaller than that in the GBM.

Region 2: $Z_{t,stat(+)} < Z_t < Z_{t,root(+)}$ the $\delta P_t$ contribution from GBM is positive, with the slope of $Kf_c(Z_t) > 0$; however the $\delta P_t$ contribution/correction from $Kf_c(Z_t)$ is still negative. This implies that $(P_{t+\delta t})_{IFBM} < (P_{t+\delta t})_{GBM}$. As $(P_{t+\delta t})_{IFBM}$ is closer to $P_t$, this peaks the returns distribution. This region is a negative feedback region as $Z_t Kf_c(Z_t)$ (product of $Z_t$ and $Kf_c(Z_t)$) is negative. The price rise in the IFBM is smaller than that in the GBM.

Region 3: $Z_t > Z_{t,root(+)}$ the $\delta P_t$ contribution from GBM is positive, with the slope of $Kf_c(Z_t) > 0$, and the $\delta P_t$ contribution/correction from $Kf_c(Z_t)$ is positive. This implies that $(P_{t+\delta t})_{IFBM} > (P_{t+\delta t})_{GBM}$. As $(P_{t+\delta t})_{IFBM}$ is further away from $P_t$, this flattens the returns distribution. This region is a positive feedback region as $Z_t Kf_c(Z_t)$ (product of $Z_t$ and $Kf_c(Z_t)$) is positive. The price rise in the IFBM is greater than that in the GBM.

Now consider Regions 4, 5, and 6 where *Z* is negative:

Region 4: $Z_{t,stat(-)} < Z_t < 0$ the $\delta P_t$ contribution from GBM is negative, slope of $Kf_c(Z_t) < 0$, and the $\delta P_t$ contribution/correction from $Kf_c(Z_t)$ is positive. This implies that $(P_{t+\delta t})_{IFBM} > (P_{t+\delta t})_{GBM}$. As $(P_{t+\delta t})_{IFBM}$ is closer to $P_t$, this peaks the returns distribution. This region is a negative feedback region as $Z_t Kf_c(Z_t)$ *(product of $Z_t$ and $Kf_c(Z_t)$)* is negative. The price fall in the IFBM is smaller than that in the GBM.

Region 5: $Z_{t,root(-)} < Z_t < Z_{t,stat(-)}$ the $\delta P_t$ contribution from GBM is negative, slope of $Kf_c(Z_t) > 0$; however the $\delta P_t$ contribution/correction from $Kf_c(Z_t)$ is positive. This implies that $(P_{t+\delta t})_{IFBM} > (P_{t+\delta t})_{GBM}$. As $(P_{t+\delta t})_{IFBM}$ is closer to $P_t$; this peaks the returns distribution. This region is a negative feedback region as $Z_t Kf_c(Z_t)$ *(product of $Z_t$ and $Kf_c(Z_t)$)* is negative. The price fall in the IFBM is smaller than that in the GBM.

Region 6: $Z_t < Z_{t,root(-)}$ the $\delta P_t$ contribution from GBM is negative; the slope of $Kf_c(Z_t) > 0$, and the $\delta P_t$ contribution/correction from $Kf_c(Z_t)$ is negative. This implies that $(P_{t+\delta t})_{IFBM} < (P_{t+\delta t})_{GBM}$. As $(P_{t+\delta t})_{IFBM}$ is further away from $P_t$; this flattens the returns distribution. This region is a positive feedback region as $Z_t Kf_c(Z_t)$ *(product of $Z_t$ and $Kf_c(Z_t)$)* $Z_t Kf(Z_t)$) is positive. The price fall in the IFBM is greater than that in the GBM.

Let us first examine the total negative feedback region corresponding to $|Z_t| < Z_{t,roots}$ (this is regions 1, 2, 4 and 5). In this feedback region, the product of $Z_t$ and $Kf_c(Z_t)$ is negative. This region models/measures the sluggishness of the market, when the agents respond in aggregate to the minimal news by reacting in an inverse manner. They are so called contrarians or irrationals. For example, when the upturn in the news is minimal, agents irrationally become impatient, sell the assets and invest in alternative products. Next, consider the two positive feedback regions where the sigmoid function has smoothly taken over and now dominates. The positive feedback region corresponding to large positive $Z_t$ values and $Kf_c(Z_t) > 0$, models/measures the agents riding the wave of euphoria, greed and irrational exuberance: "buy, buy, and buy". In the other positive feedback region, corresponding to large negative values of $Z_t$ and $Kf_c(Z_t) < 0$, the aggregate fear and panic ("sell, sell and sell") of the irrational agents is so modelled.

The negative feedback region effectively peaks the returns normal distribution near the origin; the smooth take over by the sigmoid function giving rise to the positive feedback regions flattens the distribution, whence leads to longer tails [1]. This measuring process of the irrational agent behaviour has been found to finely model the leptokurtic distribution of returns [1].

## IV. Discussion and conclusion: discovering the psychological soliton.

The Adaptive Markets Hypothesis [22] implies that the degree of market efficiency (i.e. the degree of rationality/irrationality) relates to environmental factors and that the violation of rationality is consistent with an evolutionary model of agents adapting to a consistently changing environment via simple heuristics.

In this paper, we are measuring the degree of irrationality (the degree of market non efficiency) using the stochastic function $Kf_c(Z_t)$. $K$ and $c$ are estimated for specific market data sets in specific time periods, hence are specifically and intricately linked with the degree of the environmental factors of the market and time period. However, the general shape of $Kf_c(Z_t)$ is non adaptive. It will persist through time as the markets are to some extent driven by agents who react in terms of impatience, fear and greed and will continue to do so.

Fig. 5 Three dimensional perspective of the information feedback function $Kf_c(Z_t)$ for $K = -5$ and $c = 1$ as a function of $Z_t$, and time $t$ in view of emphasizing the soliton-like aspect of the function.

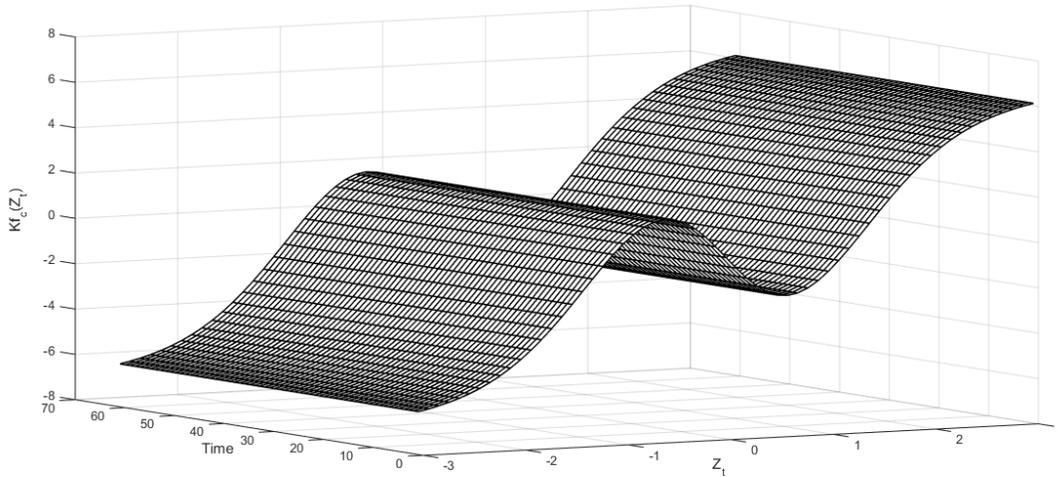

The three dimensional figure illustrating $Kf_c(Z_t)$ (for $K = -5$ and $c=1$), Fig. 5, shows that the function maintains its overall shape as it moves in time (time is represented as the third dimension). Hence $Kf_c(Z_t)$ behaves as a travelling kink into time (with a sinusoidal shape near the origin) due to the investors irrational behaviour. We can now coin that $Kf_c(Z_t)$ corresponds to a "stochastic psychological soliton" for the financial returns on a stock market. There is no need to insist that the roots positioning (dependent on $c$) and the amplitude of the kink (dependent on $K$) might be not constant, but time and case dependent as outlined above. Nevertheless, the theoretical kink characteristics (its critical $Z$ value and its amplitude) might be considered as a simple way to characterize and measure the irrational behaviour of an agent, or of a population of agents, pending fits to an overall distribution of returns in a market [1] for such distribution fits.

When analysing regularly spaced periodic time series the financial soliton will feed on this news/innovation at constant velocity. Furthermore now consider the fact that the data fed into the system can be yearly, monthly, weekly, daily or even intraday data. This data will pass through the financial soliton as discussed above, so as to model the best fit returns distribution to empirical data. We suspect that this type of modelling will then also shed knowledge on and connect this soliton structure to the multifractal behaviour of stock markets.

A warning is in order in this conclusion. One should be careful about the rigour of the mathematics and corresponding mathematical properties of solitons: soliton wave pulses are usually bell shaped (strictly categorised by *sech$^n$Z)* or travelling kinks. The present kink shape categorized by $Kf_c(Z_t)$ is a mixture of the inverse bell shaped (dark soliton) and the travelling kink shape. In essence we are mixing p.d.f with a c.d.f; i.e. a bell shape with a sigmoid function. Hence due to the shape of the soliton $Kf_c(Z_t)$, $K$ being negative, it provides the additional output that is of negative feedback when $|Z_t| < Z_{t,roots}$ and positive feedback when $|Z_t| > Z_{t,roots}$.

Financial models which incorporate stochastic volatility usually lay foundations on the GBM. However just as the GBM is appropriate to physical sciences the correct foundation for further financial modelling is the IFBM model, Eqs.(3) and (4), proposed in this paper. The IFBM both captures the leptokurtic distribution of asset returns and also provides the measure for this irrational behaviour as captured by the psychological soliton.

Notice in fine that Engle [30] wrote that "the use of exogenous variables to explain changes in variance is usually not appropriate"; we insist that there is no exogenous variable indeed in [1] nor here above.

For ending this conclusion, we can suggest a few questions for further work, expanding the present findings. For example, one can ask

1. What is the differential equation whose solution is $Kf_c(Z)$ ?
2. What is the discrete form financial econometrics version of Eq. (3), as Eq. (1) is the discrete version of Eq. (2) ?
3. Is the IFBM model leading to a better volatility clustering description ?
4. Should other non linear analytic functions that Eq. (5) be considered, and why ?
5. Can one further consider extensions of the IFBM (i) toward volume distributions, (ii) technical analysis considerations, or (iii) examining soliton-soliton, soliton-antisoliton interactions, on other spaces, (e.g., time, wave vector, and energy [31]), and (iv) in the vicinity of financial crashes ?

6. Can the IFBM modelling approach resolve anomalies and doubts on results of EMH tests [32] in financial markets ?
7. Is the information feedback function Eq. (5) fully and sufficiently explaining the empirical evidence of the psychological effect discussed in [33], - which leads to underestimation of small changes and overreaction to big changes across the worldwide stock markets ?